\documentclass[reprint,https://www.overleaf.com/5940909bvkbhc amsmath,amssymb,showpacs, aps,prd]{revtex4-1}
\UseRawInputEncoding
\usepackage{amsmath}
\usepackage{tikz}
\usepackage{amsfonts}
\usepackage{amssymb}
\usepackage{mathrsfs}
\usepackage{latexsym}
\usepackage{footnote}
\usepackage{hyperref}
\usepackage[english]{babel}
\usepackage{cmap}
\usepackage[T1]{fontenc}
\usepackage{textcomp}
\usepackage{setspace}
\usepackage{graphicx}
\usepackage{dcolumn}
\usepackage{bm}
\usepackage{float}
\usepackage[short]{optidef}
\usepackage[text={16cm,24cm}]{geometry}
\usepackage{mathptmx}
\usepackage[compact]{titlesec}
\titlespacing{\section}{0pt}{2ex}{1ex}
\titlespacing{\subsection}{0pt}{1ex}{0ex}
\titlespacing{\subsubsection}{0pt}{0.5ex}{0ex}
\begin{document}

\preprint{APS/123-QED}

\title{ Searching signature of neutrino-nucleus coherent scattering with M\"ossbauer Spectroscopy }

\author{$^1$$^,$$^2$C. Marques}
\author{$^1$G. S. Dias}
\author{$^3$H. S. Chavez}
\author{$^2$S. B. Duarte}
\affiliation{$^1$Department of Physics - IFES - Instituto Federal do Esp\'{i}rito Santo (IFES), \\ Av. Vit\'{o}ria 1729, Jucutuquara, Vit\'{o}ria, ES, Brazil, CEP 29040-780}
\affiliation{$^2$Centro Brasileiro de Pesquisas F\'{i}sicas (CBPF) Rua Dr. Xavier \\ Sigaud 150, Urca, RJ, Brazil, CEP 22290-180} 
\affiliation{$^3$UCL-Faculdade do Centro Leste, Rodovia ES 010, Km 6 \\ BR 101, Serra-ES, Brazil}

\email{celiom@ifes.edu.br}
\email{gilmar@ifes.edu.br}
\email{helderch@hotmail.com}
\email{sbd.cbpf.rj@gmail.com}
\date{\today}
\begin{abstract}
The M\"ossbauer spectroscopy is presented as an alternative experimental technique to be pursued in the detection of Coherent Elastic $\nu$-Nucleus Scattering (CENNS). The neutrino transferred energy in the neutrino-nucleus interaction causes a perturbation at the nuclear levels which are responsible for gamma radiation in M\"ossbauer resonance. The main characteristic of M\"ossbauer spectroscopy is the recoil less condition imposed on the nuclei imposed by quantum restrictions of the crystalline array potential. 
Coherent neutrino scattering is expected to occurs low energy regime of the beam, typically below $30$ MeV, at higher neutrino energies small incoherence component appears in the scattering\cite{bednyakov} as disturbances within the internal nuclear degrees of freedom.
In this work we have shown that using anti neutrino reactor beam, with the accuracy of M\"ossbauer technique its possible to obtain a measurable signature of the Coherent neutrino Scattering.   



\end{abstract}


\maketitle

                
\section{\label{sec:level1}Introduction}
A direct evidence of the Coherent Elastic $\nu$-Nucleus Scattering (CENNS) has been considered an experimentally challenging task to better understand weak neutral current in the context of the Standard Model of Particle Physics(SM) \cite{freedman,giunt,brice}. An evidence of the process  has been intensively pursued by many experimental collaborations in the last four decades~\cite{drukier,formaggio,giunt,brice,collar} developing a great effort to the direct detection of this weak process, which has the largest predicted cross section in the low energy range for neutrinos, typically  below $50$ MeV the coherence condition is matched. 

Recently, the COHERENT collaboration report the first undoubtedly measurement of the CENNS process\cite{akimov}. An efficient scintillator detector was used in the experiment carried out at the  Oak Ridge National Laboratory. With a Spallation Neutron Source it was produced an extremely intense neutron beam, which was scattered by a mercury target generating a secondary pion beam. Produced pions decays into an intense  neutrino flux ($\approx 10^{11}/$s) with energy in the range of $16$ to $53$~MeV~\cite{akimov}. The pulsed neutrino flux was scattered by $14.6$ kg crystal made of CsI doped with Sodium atoms. The experimental setup was properly structured to prevent any contamination from external sources of neutrons and neutrinos, like atmospheric or solar and galactic neutrinos. This experiment accumulate CENNS events during fifteen months produced in accordance to the SM prediction.

Here we are searching for a signature of the CENNS using M\"ossbauer nuclear spectroscopy(MS) \cite{greenwood,gruverman}. The great domain of the technique all over the world makes possible a broad experimental setups which could be developed in relatively low cost profiles.

As it is well known the main characteristic of M\"ossbauer technique is the recoil less interaction of the electromagnetic radiation with the nucleus in a crystal sample. This aspect is assumed to be preserved in the CENNS process with absorber nucleus used in M\"ossbauer measurement. The nucleus maintain its fixed position in the local minimum of the crystal potential lattice. The effect of the energy transferred by neutrinos to the nucleus is to induce a perturbation in the inner degree of nuclear structure. We assume that the quantum state of the valence neutron is slightly modified. Consequently, the nuclear volume is changed and the consequent change in the isomeric shift can be observed with the typical accuracy of this spectroscopic technique. To determine our prediction, we assume that only the valence neutron is perturbed at the structure of a characteristic nucleus of  M\"ossbauer machine (e.g. a typical one is  $^{57}$Fe). With this transition between single particle states of the valence neutron  in the CENNS process, we calculate the change in isomeric shift.  Pauli blocking prevent transitions at the inner structure of the nuclear core. 
 
In next section of this letter, we summarize the main characteristics of CENNS. Our proposal of use  M\"ossbauer technique to observe CENNS is presented in more details in section III. The isomeric shift correction due to CENNS process is calculated in section IV. In Section V it is estimated the event rates for some M\"ossbauer isotopes used as absorber. Section VI presents our main conclusions.
                  
\section{The Main Characteristic of CENNS}

 The CENNS was proposed theoretically by Freedman~\cite{freedman}in 1974. A Feynman diagram of this weak process is shown in Fig.\ref{diagram}. The effective Lagrangian to describe the process is given by 
\begin{equation}
L = G_{F} L^{\mu}J_{\mu},
\end{equation}
where $G_{F}$ is the Fermi constant, $L^{\mu}$ the lepton current, and $J_{\mu}$ is the hadron current inside the nucleus. 
Experimental efforts have been developed in the detection of CENNS, some of them represented by large scientific collaborations namely, COHERENT \cite{coherent,akimov}, CONNIE \cite{connie} and TEXONO \cite{tex}, among others. As mentioned before, after decades of searching only recently the COHERENT Collaboration~\cite{akimov} announced the first irrefutable detection of CENNS. Here we are interested in discuss another strategy in order to see this interactions which don't competes to another experimental plants, but add to them one more tool to seek properties about neutrinos physics and possibly searchs of beyond Standard Model Physics(BSM), as are now considered by others\cite{gratta,radeon,bednyakov}.

It is well known that the coherence aspect of CENNS, requires $qR \ll 1$, with $q$ being the transferred momentum and $R$ the nuclear radius. This implies that the wavelength of neutrinos will be comparable to the nuclear radius. Detailed discussions about the phenomena can be found in Refs.~\cite{freedman,mosel,krauss,formaggio,brice} and references therein. We stress the fact that the cross section of this process has the largest value $(\sigma \approx 10^{-38}$ cm$^{2})$ at least four orders of magnitude larger than other neutrino interactions in the same low-energy regime~\cite{formaggio}.

 \begin{figure}[H]
 \begin{center}
 \includegraphics[scale=0.3]{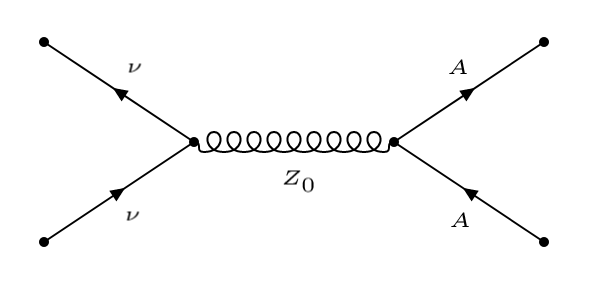}
 \caption{Feymann diagram of the  CENNS process.}
 \label{diagram}
 \end{center}
 \end{figure}

The Freedman differential cross section for this process is~\cite{freedman,giunt,kate}
\begin{equation}
\frac{d\sigma_{\rm CENNS}}{dT}=\frac{G_{F}^{2}}{4\pi}Q_{w}^{2}M_{A}F^{2}(q^{2})\bigg(1 -\frac{T}{E_{\nu}}- \frac{M_{A}T}{2E_{\nu}^{2}}\bigg) ,
\label{cross} 
\end{equation}
where $T$ is the transferred energy to the nucleus, ${A}$ is the target mass, $E_{\nu}$ is the neutrino energy and $Q_{w}= N -Z(1-4$sin$^{2}\theta_{w})$ is the weak charge, which depends on the number of neutrons ($N$) and protons ($Z$). Here $\theta_{w}$ is the Weinberg angle satisfying sin$^{2}\theta_{w}\approx 1/4$ and the proton contribution is negligible. The last fact made CENNS a very sensitive probe to nuclear neutron density~\cite{balant}. The form factor $F(q^{2})\to 1$ as $q\to 0$ defines the coherence condition and so the elastic scattering at nucleus\cite{bednyakov}. 

We remind that as a weak process, the interaction involved in the CENNS should be many order of magnitude greater than the gravitational phenomena. Even so, the MS was successfully employed to measure the gravitational red shift of light by Pound and Rebka \cite{pound,snider} at the end of the fifties. Thus, we hopefully expect that a properly MS array could be used in the study of eletroweak phenomena.                       

\section{The M\"ossbauer Technique Applied to Detect CENNS} 

One of the main characteristics of the MS is that the nuclei in the absorber material of the experiment  are recoil less when interacting with gamma photons coming from the source parent nucleus decaying. This condition is fundamental for the resonant absorption radiation in the MS. We point out here as a novel assumption, the recoil less feature also to the CENNS. This assumption is compatible with the fact that the transferred energy to the nucleus are in the same scale of the MS gamma photon energy. Thus the hyperfine levels can be accessed or perturbed $Z^{0}$ exchange at the CENNS. The transferred energy-momentum by the $Z^{0}$ is assumed to be transmitted to the valence neutron of MS isotope, slightly modifying the nuclear mean field to the valence neutron. The wave function and energy levels are perturbed promoting a typical isomeric shift correction in the MS spectrum.
Considering the discussion in reference\cite{bednyakov}, this implies that the perturbed M\"ossbauer nuclides by the CENNS, open the incoherent channel in the sense of the nuclear levels.
The recoil less condition is satisfied for both processes, the MS photon resonance and for the coherent neutrino scattering. 

In fact, the fraction $f$ of the recoil less nuclei in $Z^{0}$ exchange between neutrino and nuclei in the CENNS can be analyzed similarly to the case of the  gamma radiation interaction. It can be shown that the fraction of recoil less events can be put in the form of Debye-Waller factor~\cite{marf}, which, for the CENNS, takes the form
\begin{equation}
f=\exp{\left(-\frac{T^{2}}{Mc^{2}\hbar\omega}\right)} ,
\end{equation}
where $T=E_{\nu}^{2}/2Mc^{2}$ is the energy transferred by the $Z^{0}$ to target nucleus of mass $M$. Here $\hbar\omega \approx 10^{-3} $ eV for Fe, Co etc, is the order of magnitude of energy lattice vibrations. In the range of neutrino energies below $\approx 50$~MeV the recoil less $f$ factor is essentially unity. Therefore, we argue that this small energy fraction will be accommodated inside the nucleus in a form of a suitable enhancement of the isomeric displacement. 
                        
\section{Isomeric Shift Correction due to the CENNS Interaction}

When a M\"ossbauer nucleus absorbs a KeV gamma radiation photon, this could be absorbed in some transition levels, which is defined by the hyperfine interactions of that nucleus surroundings. In CENNS interaction however, the KeV energy is transferred by $Z^{0}$ to the nuclear mean field, and due the weak charge of proton be essentially zero, the neutron levels are changed absorbing the energy momentum transferred. 


The unperturbed $^{57}$Fe valence neutron is at a state of definite angular momentum, given by the common distribution of the neutron and proton content \cite{weiss} in nuclear shell model -- its wave function is regular at the origin (typically a Bessel function). 
The effect of the nuclear extended charge instead of punctual nucleus in the Coulomb contribution in the nuclear levels could be written as:
\begin{equation}
\delta I_{s}=\frac{4\pi Ze^{2}R^{2}}{5}\big(\frac{R_{exc}-R_{gs}}{R}\big)(\psi_{l=0}^{2}(0)_{a}-\psi_{l=0}^{2}(0)_{s})
\label{deltas}
\end{equation}.

Where $Z$ is the number of protons of nucleus, $R$ is the mean radius of the charge distribution and $R_{exc/gs}$ is the excited and ground state radius of nucleus. In the literature the difference $(R_{exc}-R_{gs})$ is reported as of order $-10^{-3}R$ \cite{marf}. In our estimate here, we consider $R=1.3A^{1/3} fm$, $A$ being the mass number of nuclei in consideration. The difference between the $\psi's$ at $r=0$ measures the deviation of the probability that the electron s -$(l=0)$- could be find at nuclear volume of the source(s) and in the absorber nucleus(a).

In fact, we are considering high Z nucleus like $Fe^{56}$  where the QED corrections, specially, electron polarizability of vacuum diminishes the s electron  binding energy\cite{kuchiev}.

This effect was noted long ago by Lamb and Retherford. The $2s_{1/2}$ and $2p_{1/2}$ has small differences in their binding energies, giving rise to the well known Lamb Shift. 

The Lamb shift correction to electronic levels is well calculated in the literature and is given as:

\begin{equation}
\delta E_{L} = \frac{\alpha}{\pi}\frac{(Z\alpha)^{4}}{n^{3}}F(Z\alpha)m_{e}c^{2}
\end{equation}.

Where $\alpha$ is the fine structure constant, $n$ is the principal quantum number, and Z is the number of protons at nucleus. $F(Z\alpha)$ is in general a small varying function. So the important thing is that the Lamb correction scales approximately as $Z^{4}/n^{3}$ and is largest for the 1s electrons. 

This fact, connects the Lamb Shift in electronic energy levels and the isomeric displacement $\delta I_{s}$ - see \ref{deltas} - of the nuclear levels,  which  originates from the s electrons finite probability in the nuclear volume. 
Solving the Dirac equation for the central potential $Ze^{2}/r$ and taking the positive energy solutions, we have\cite{kuchiev}:
\begin{equation}
\psi_{n,l+1/2}(r)=\left( \begin{array}{c}
g_{n,l}(r)Y_{l}(\Omega)  \\ 
if_{n,l}(r)Y_{l}(\Omega)
\end{array} \right)
\end{equation}
normalization requires 
that integration of $g^{2}+f^{2}$ from $r=0$ to infinity  gives one. Therefore, considering the finite size of the nucleus, the probability that the electrons is found inside the nuclear volume is:
\begin{equation}
B^{2}_{n,l}(R)=\frac{1}{4\pi(m_{e}c^{2})^{3}}\bigg(\frac{\hbar c}{a_{0}}\bigg)^{3}\bigg(\frac{a_{0}}{R} \bigg)^{2}[g_{n,l}^{2}(R)+f_{n,l}^{2}(R) ]
\label{prob}
\end{equation}.
Where $a_{0}$ is the Bohr radius of Hydrogen and $R$ is the nuclear radius. 
When the neutrino exchanges a virtual $Z_{0}$ we are assuming here that the neutrons and protons perceives differently this interaction because the  crystalline potential. In fact if we perform a phenomenological analyses in the point of view of the nucleus, the Hamiltonian description can be cast in form:

\begin{equation}
H_{N}=\sum_{j}^{A}\gamma^{0}M_{j}c^{2}-i\hbar\gamma^{i}\partial_{i}+V_{eff}+ \frac{1}{4}F_{\mu\nu}F^{\mu\nu} -(G_{F}L^{\mu}J_{\mu}+ 
\scriptsize
\frac{1}{c}\int A^{\mu}J_{\mu})
\end{equation}.

This could be  though as the unperturbed (Dirac + Maxwell) part, and the interacting part, that involves a Weak plus the one involving the eletromagnetic field perceived by the recoil less nucleus after the $Z_{0}$ exchange. The last term in the Hamiltonian, clearly distinguishes protons against neutrons. If we make a static approximation at the Hamiltonian, the second part of the interacting Hamiltonian can be seen as:
\begin{equation}
\frac{1}{c}\int A^{\mu}J_{\mu}\Rightarrow V_{c} \rho_{Z}
\end{equation}
Where, $V_{c}$ is the potential of the crystalline array and $\rho_{Z}$ is the proton density of the nucleus. 
The Weak part is basically non zero due the neutron density of the nucleus as we consider the Weak charge of the protons  almost zero. We follow \cite{giunt,bednyakov} and stress that some decoherence will be present in the nuclear system and we have this as a $\frac{\delta R}{R}$ which appear due to the disturbing at the ground state wave function of the valence neutron. The variation of the  quadratic mean radius of the nucleus leads to:

\begin{equation}
\frac{\delta R}{R}\bigg\vert_{Z_{o}}=\sqrt{2}\frac{197}{1.3A^{5/6}\sqrt{\hbar \omega m_{N}c^{2}}}
\label{deltaz}
\end{equation}.

Where $M_{N}$ is the neutron mass and for estimations $\hbar\omega$ is of order $0.1$ KeV or less.
Considering this small modification of the nucleus we can expand \ref{prob}:

\begin{equation}
B^{*2}_{n,k}(R+\delta R)=B^{2}_{n,k}(R)+\frac{\partial B^{2}_{n,k}(R)}{\partial R}\delta R + ... 
\end{equation}.

Where star(*) means after $Z_{0}$ exchange. Plugging the obtained result in \ref{deltas} we have the following expression to the modified Isomeric dislocation:

\begin{eqnarray}
\nonumber
&
\delta I_{s}^{*}=\delta I_{s}- \frac{3Z^{4}e^{2}}{10\pi}\frac{\delta R}{R}\bigg\vert_{\gamma}\frac{\hbar^{3}c^{3}}{(m_{e}c^{2})^{3}a_{0}^{4}}\frac{\delta R}{R}\bigg\vert_{Z_{o}}
\end{eqnarray}. 

In fact this result depends on the relative variation $\frac{\delta R}{R}\bigg\vert_{Z_{o}}$, caused in the nuclear radius by the virtual $Z_{0}$ exchanged with neutrino current. Assuming the expression $(8)$ and that $\hbar\omega$ is of order of keV, the $\delta I_{s}^{*}$ presents $ ~ 2.41*10^{-7}eV$ difference to that without nuclear recoil. We observe that this small energy contribution to this level asymmetry between font and absorber nuclei, scales with $Z^{4}$ as we see for the Lamb Shift observed in the  electronic levels. This tells us that the effective treatment of $Z_{0}$ interaction with nucleus, could emulate the vacuum polarization for the weak field which is perceived by the proton and neutron array inside the nucleus.  

Considering the values of constants in the expression, and a nucleus with $A~60$, we can obtain the correction at Isomeric displacement of order $0.25  \mu eV$.

The M\"ossbaur Spectroscopy(MS) has the capability of resolves differences in hyperfine energies of order $10^{-10} eV$, so we appoint here that, if we take a MS near($10m$ away) the reactor font, and other 
identical measure at reactor off situation, we can see after subtraction of the resulting spectra, some experimental points not coincident, which can reveal this contribution of the neutrinos. 


Another point of view that gives the same results is the perturbed treatment. We can define a $\lambda$ parameter that is associated to the ratio between the square of the transferred energy and the energy difference of the non-perturbed energy states~\cite{kuchiev} of the valence neutron. We have the equation below:
\begin{equation}
\lambda = \frac{3E_{\nu}^{2}}{8Mc^{2}(E_{5/2}-E_{3/2})}. 
\label{lambda}
\end{equation}
The term $(E_{5/2}-E_{3/2})$ is the difference between the energy of the non-perturbed states of the valence neutron. In $^{57}$Fe case, this is responsible for the emission of $14.4$ KeV photon which is emitted and absorbed resonantly without nuclear recoil.

In the context of the shell model for Woods-Saxon potential with spin orbit term\cite{samuel,weiss,charm,kuksa}, the two states of valence neutron for the $^{57}$Fe can be described by spherical Bessel functions, $j_{3/2}(kr)$ and $j_{5/2}(kr)$. In this case we consider that the neutron field which absorbs the $Z^{0}$, comes to be described by a superposition, behaving like $ j_{3/2}(kr)+ \lambda j_{5/2}(kr) $. We have used for the wave number of the valence neutrons $k \approx 0.5$ fm$^{-1}$, which is, as usually, approximately the inverse of twice neutron radius. With this approach  the isomeric shift can be calculated~\cite{greenwood} as being

\begin{equation}
\delta I^{*}_{s}=\frac{4\pi Ze^{2}R_{gs}^{2}}{5}\Bigl(\frac{R_{exc}-R_{gs}}{R_{gs}}\Bigr)[\psi_{l=0}^{2}(0)_{a}-\psi_{l=0}^{2}(0)_{s}] ,
\label{deltas}
\end{equation}

where $Z$ is the number of protons in the nucleus, $R_{gs}$ is the mean radius of the charge distribution at the ground state of nucleus and $R_{exc}$ for the $Z^{0}$ excited nuclear radius, respectively, the $\psi 's$ are the $s$ electrons wave functions, evaluated at the origin~\cite{marf,greenwood} for the absorber and the source of gamma radiation. In the literature, the difference between nuclear radius of the excited nucleus and the ground state is $(R_{exc}-R_{gs})$, which in conventional gamma resonance is reported to  be $\approx 10^{-3}R_{gs}$ \cite{marf,greenwood}. Our estimate for $\frac{R_{exc}-R_{gs}}{R_{gs}}$, calculated using the perturbed neutron wave function is given by $1/r\sqrt{1+\lambda^{2}}(j_{3/2}(kr)+ \lambda j_{5/2}(kr)) $ above is 
$\approx 10^{-4}$. The two level description taking the eletroweak as non diagonal terms of the nuclear unperturbed system, was still analysed, but more profound analyses in this direction will postponed to future works. By now we stress with this result and Eq. \ref{deltas}, we can obtain the correction at isomeric shift induced by the weak process ($Z^{0}$ exchange) in CENNS. As we can see only one order of magnitude smaller than the typical  characteristic $\gamma$ measurements. This value for $\delta_{I^{*}_{s}}$ is perfectly solved with the MS technique accuracy, namely $10^{-10}$
eV \cite{pound,snider,greenwood}.  Consequently, we point out that if we take subtraction of a MS measurement without the neutrino flux and other result of identical measure with the reactor neutrino beam, we would reveal the contribution of the CENNS interactions.
           
\section{Event Rates for M\"ossbauer Isotopes}

In this section we estimate the  expected event rate $n$, for some nuclei of interest in M\"ossbauer Spectroscopy. This number could be given by the expression \ref{nevi}.  

\begin{equation}
n = N_{t}\int_{E_{min}}^{E_{max}}dE_{\nu}\Phi(E_{\nu})\frac{d\sigma(E_{\nu})}{dE_{\nu}},
\label{nevi}
\end{equation}

Where $N_{t}$ is the number of target nuclei, which we assume of order $10^{23}$, $E_{\nu}$ is the energy of neutrinos at incident flux, $\Phi$ is the reactor flux of neutrinos, $\sigma$ is the cross section of CENNS. $E_{max/min}$ is the maximum/minimum energy at the spectrum of neutrinos. In the case of M\"oessbauer nuclei, the energy exchange of Z and the target nucleus is integrally absorbed by the neutron field, and the overall nucleus has no momentum inside the crystal lattice. Then the cross section is maximal and given by:
 
\begin{equation}
\sigma_{cenns}=\frac{G_{F}^{2}E_{\nu}^{2}N^{2}}{4\pi},
\label{nev}
\end{equation}

In this expression $G_{F}$ is the Fermi constant and $N$
is the neutron number, $E_{\nu}$ is the energy of neutrinos. The proton contents contribution is ignored due its tiny weak charge. The energy distribution of the neutrino flux from the reactor neutrinos is not well established in complete  details\cite{kopeikin}. However some estimations can be used in order to have a  practical definition of the rate and number of events in an experimental test. 

The reactor neutrino flux is not well theoretically understood in its full extent. As we see at the figure 2 below, it presents a bump of anti neutrinos at $4.5$ MeV, which is not yet well explained. We know that  it holds a very high quantity of neutrinos/$cm^{2}s$, of order $10^{13}$.  It ranges between $0$ to $E_{max}=10$ MeV.
 
 \begin{figure}[htb]
 \begin{center}
 \includegraphics[scale=0.53]{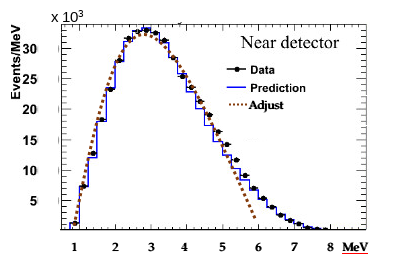}
 \caption{ The expected
anti neutrino spectrum weighted by the inverse beta decay(IBD) cross
section from Daya Bay(black points)\cite{dayb}. Predicted and measured prompt energy spectra(blue). Our adjust(red points) taking to account only the principal region of spectra in order to obtain event rate. The point values of figure was extracted from \cite{vorobel}.}
 \label{diagram}
 \end{center}
 \end{figure}
 
In our estimation we use the normalized function:

\begin{equation}
    \Phi(E_{\nu})=22.6*10^{-3}(E_{\nu}-\frac{1}{6}\cdot10^{-6}E_{\nu}^{2})
\end{equation}

 as the mostly simple curve that fits approximately the spectrum for the reactor showed at figure $2$. With these assumptions, the event rates $n$ for some M\"ossbauer isotopes are displayed in the graph below and more specifically for three MS isotopes at table $1$. \\

\begin{table}[htb!]
\centering
\caption{Event Rate for possible M\"ossbauer isotopes calculate using Eq. \ref{nevi}}
\vspace{0.1cm}
\begin{tabular}{||p{3.5cm}||p{3.5cm}||} 
     \hline
  $10^{23}$ Nucleus of: &  Respective Event rate: \\ [0.5ex] 
 \hline\hline
   $^{26}$Fe & $\approx 10^{2}/day$ \\ 
$^{45}$Rh & $ \approx 10^{3}/day$ \\
 $^{97}$Bk  & $ \approx 10^{3}/day$ \\ [1ex] 
 \hline
 \end{tabular}
\label{table:1}
\end{table}
\begin{figure}[htb]
 \begin{center}
 \includegraphics[scale=0.63]{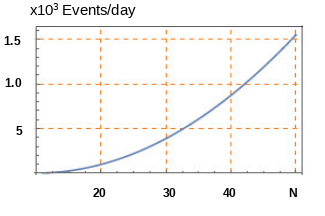}
 \caption{Events calculated from expression $8$. We are considering $\approx 10^{23}$ nucleus subjected to an integrated reactor anti neutrino flux of order $10^{13}/cm^{2}s$}
 \label{diagram}
 \end{center}
 \end{figure}

The isotopes Bk and Rh present higher event rates than the Fe, however there are some difficulties in CENNS observation in the use of them. The separation of the valence levels energy, are some hundreds of keV, making the parameter $\lambda$ in Eqs.$(4-5)$ almost null. Additionally the measure of the spectra of these elements is hard to calculate, due to their tiny relative abundance. This appoints the conventional Iron isotope as a favorable choice of the  CENNS experiment using the M\" ossbauer Technique.

\vspace{0.7cm}           
\section{Conclusions}

This work develops model calculation to determine some estimate of the rate and number of events to give support for the use of MS spectroscopy as a suitable technique to see the CENNS process.

The isomeric shift correction obtained when the machine is exposed to a neutrino flux of reactors is the signature of  contribution at the MS experiment. Our estimate is a correction of the order $\sim 10^{-7}$~eV for a system using  $^{57}$Fe isotope, this value is greater than the typical energy resolution of this technique,$\sim 10^{-10}$~eV. We then guess, that future MS experiments could be suitable to integrate the neutrino experimental plants. 
\vspace{.10cm}

\begin{acknowledgments}
S. B. Duarte acknowledge financial support from CNPq. C\'elio Marques are grateful to Pedro Cavalcanti Malta, Gustavo Pazzini de Brito, Jos\'e A. Helay\"el-Neto, H\'elio da Motta and Arthur M. K\'os of CBPF-Brazil for the encouragement and discussions during the development of this work also to Dmitry Naumov and Maxim Gonchar of Joint Institute for Nuclear Research due encouragement and help to understand better the reactor anti neutrino flux and information about flux of Daya Bay experiment.
\end{acknowledgments}

\end{document}